# Three-dimensional Reconstruction of Coronal Mass Ejections by CORAR Technique through Different Stereoscopic Angle of STEREO Twin Spacecraft

Shaoyu Lyu[23], Yuming Wang[1234], Xiaolei Li[23], Jingnan Guo[23], Chuanbing Wang[23], and Quanhao Zhang[23]

**Abstract**

Recently, we developed the Correlation-Aided Reconstruction (CORAR) method to reconstruct solar wind inhomogeneous structures, or transients, using dual-view white-light images (Li et al. 2020; Li et al. 2018). This method is proved to be useful for studying the morphological and dynamical properties of transients like blobs and coronal mass ejection (CME), but the accuracy of reconstruction may be affected by the separation angle between the two spacecraft (Lyu et al. 2020). Based on the dual-view CME events from the Heliospheric Imager CME Join Catalogue (HIJoinCAT) in the HELCATS (Heliospheric Cataloguing, Analysis and Techniques Service) project, we study the quality of the CME reconstruction by the CORAR method under different STEREO stereoscopic angles. We find that when the separation angle of spacecraft is around 150°, most CME events can be well reconstructed. If the collinear effect is considered, the optimal separation angle should locate between 120° and 150°. Compared with the CME direction given in the Heliospheric Imager Geometrical Catalogue (HIGeoCAT) from HELCATS, the CME parameters obtained by the CORAR method are reasonable. However, the CORAR-obtained directions have deviations towards the meridian plane in longitude, and towards the equatorial plane in latitude. An empirical formula is proposed to correct these deviations. This study provides the basis for the spacecraft configuration of our recently proposed Solar Ring mission concept (Wang et al. 2020b).

**Keywords**: Solar coronal mass ejections; Solar wind; Heliosphere; Optical observation

## 1. Introduction

Since the 1990s, many spacecraft have been launched for observing various solar phenomena. One of the most concerned solar eruptions is the coronal mass ejection (CME), which injects a large amount of magnetized plasma from the corona into heliosphere and creates major disturbances in the interplanetary medium. CMEs propagating towards the Earth may couple with the Earth's magnetosphere and trigger magnetic storms (Gosling et al. 1990), bringing about troubles in the operation of space-borne and ground-based instruments, causing potential problems for modern communications. To better understand CMEs, instruments for in situ measurements of their interplanetary signatures and remote sensing observations of their origins at/near

[1] Corresponding author, ymwang@ustc.edu.cn

[2] CAS Key Laboratory of Geospace Environment, Department of Geophysics and Planetary Sciences, University of Science and Technology of China, Hefei, China

[3] CAS Center for Excellence in Comparative Planetology, University of Science and Technology of China, Hefei, China

[4] Mengcheng National Geophysical Observatory, University of Science and Technology of China, Mengcheng, China

the Sun are required. In particular, white-light observations by coronagraphs and heliospheric cameras are essential to study the morphological and dynamical properties of CMEs in three-dimensional (3D) space, e.g., the Large Angle and Spectrometric Coronagraph (LASCO, Brueckner et al. 1995) on board Solar and Heliospheric Observatory (SOHO, Domingo et al. 1995), the Solar Mass Ejection Imager (SMEI, Eyles et al. 2003), the coronagraphs (COR-1 and COR-2) and heliospheric imagers (HI-1 and HI-2, Harrison et al. 2005) in the SECCHI suite (Howard et al. 2008) on board the Solar Terrestrial Relations Observatory (STEREO, Kaiser et al. 2008), the Wide-field Imager for Solar Probe (WISPR) on board the Parker Solar Probe (PSP, Fox et al. 2016), and the coronagraph METIS and heliospheric imager SoloHi on board the Solar Orbiter (Muller et al. 2013).

Based on multi-view observations, the propagating directions and velocities of solar wind transients can be calculated by comparing the tracks on the time-elongation profiles, i.e., the J-maps (Sheeley et al. 1999), from different vantage points. For CMEs with sophisticated structures, some geometrical models with different assumptions are developed to obtain the kinematic parameters of CMEs. Such forward modelling (FM) techniques include the cone or ice-cream cone models (Xie et al. 2004; Xue et al. 2005; Zhao 2008; Zhao et al. 2002), the self-similar expansion model (Davies et al. 2012; Davies et al. 2013; Volpes & Bothmer 2015; Wang et al. 2013), and especially the Graduated Cylindrical Shell (GCS) model (Thernisien 2011; Thernisien et al. 2009; Thernisien, et al. 2006). Compared with these modelling methods, the Geometric Localization method (de Koning et al. 2009; Pizzo & Biesecker 2004) can reproduce the 3D shapes of CME by triangulating the boundary of imaging CMEs without a prior model. It was further improved into the 'Mask Fitting' method (Feng et al. 2013; Feng et al. 2012) which corrects the CME shapes in accordance with images from three different points. For single-perspective polarized Thomson-scattering images, the polarization ratio technique (Dere et al. 2005; Moran & Davila 2004; Moran et al. 2010; Susino et al. 2014) can locate the center of mass along the lines of sight and thus generate a density distribution. Multi-view observations may improve the accuracy of the results from the polarimetric method. Considering the correlation between the same CME patterns from different vantage points, the Local Correlation Tracking method (Feng, et al. 2013; Mierla et al. 2010; Mierla et al. 2009) calculates the correlation coefficients between images from different perspectives to determine the 3D structure representing CMEs. Recently, Li et al. (2018, 2020) successfully developed a so-called CORrelation-Aided Reconstruction (CORAR) method to recognize and determine the location, angular size and propagating direction of solar wind transients in 3D space by using STEREO HI-1 images from two perspectives.

Different methods based on multi-view observations may require different optimal stereoscopic angles for best results. de Koning, et al. (2009) found that the Geometric Localization method works best when the separation angle of spacecraft is between 30° and 150°. For tie-pointing technique, the reconstruction error is inversely correlated with the base angle between two STEREO spacecraft (Mierla, et al. 2010; Mierla et al.

2011). Liewer et al. (2011) also studied the Tie-pointing and Triangulation method (T&T), which gives reliable results on the CME propagation direction when the stereoscopic angle is within 50°. The Local Correlation Tracking plus Triangulation (LCT-TR) method works well for small separation angles (Mierla, et al. 2009), but does a poor job for large separation angles (Feng, et al. 2013). Our previous work studied the three-dimensional reconstruction of small-scale transients by the CORAR method with simulated blobs, and concluded that a dual-spacecraft angle of about 120° is the most suitable scheme for the CORAR technique (Lyu, et al. 2020). However, different from blobs, CMEs are large-scale structures. It should be studied if the best separation angle of the two spacecraft for more complicated CMEs is the same or similar with our previous work. Thus, in this paper, we apply the CORAR technique to the observed CMEs by HI-1 to achieve the optimal stereoscopic angle for large-scale heliospheric transients. In Section 2, we introduce the CORAR method and HI-1 data for reconstruction, as well as the classification assessment of reconstruction quality and collinear effect. Section 3 analyzes the goodness of reconstruction in different angular intervals to achieve the optimal stereoscopic angle, and derives the deviations or errors from CME propagating directions, which are further discussed in Section 4. Finally, we provide the conclusion in Section 5.

## 2. Data and Method

We use the CORAR method to process the HI-1 Level-2 white-light images from December 2008 to February 2012 for 3D reconstruction. During the interval, the two spacecraft STEREO-A and B moved away from each other at the speed of about $22°$ per year, with their separation angle (the angle between two spacecraft bisected approximately by the Sun-Earth line) increasing from about $85°$ to $225°$. The field of view (FOV) from HI-1 cameras is 20°×20°, observing the area from 4° to 24° outward from the Sun. The pixel resolution of HI-1 images is 1024×1024, and the time interval between two successive images is 40 min, showing the evolution of fine structures in solar wind transients. Before reconstruction, HI-1 images are processed by removing star light and noise, and the pixel shifts of continuous images are corrected for running-difference.

Compared with other manual approaches for locating CME features, the CORAR method can automatically recognize the patterns belonging to the same transient in the dual-view images. After data preprocessing, the selected images in the same time period are projected on 81 meridian planes from -80° to 80° in longitude of the Heliocentric Earth Ecliptic (HEE) Coordinate, with the grid resolution of 1° in latitude from -80° to 80° and 0.4 solar radii ($R_\odot$) in radial direction. Then, the program calculates the distribution of cc, which is the Pearson Correlation Coefficient between the projections of dual images and further corrected by the local signal-noise ratios, to present the three-dimensional structures of CMEs. The size of the sampling box searching for CME patterns during cc calculation is 11×21×5, representing 10° in latitude, $8R_\odot$ in radial distance and 160 min with 5 time steps, respectively. For saving storage, we only store effective cc data with values higher than the threshold of 0.5, which is considered as the high-cc region to match solar wind transients. More details of the CORAR process can be seen in Li, et al. (2020).

We use the CMEs listed in the Heliospheric Imager CME Join Catalogue (HIJoinCAT), as well as their kinematic properties in the Heliospheric Imager Geometrical Catalogue (HIGeoCAT, Barnes et al. 2019), to achieve our goal in this study. These two catalogues are both generated by the Heliospheric Cataloguing, Analysis and Techniques (HELCATS, Barnes, et al. 2019; Harrison et al. 2018; Murray et al. 2018; Plotnikov et al. 2016; Pluta et al. 2019) project funded under the European Union's Seventh Framework Programme for Research and Technological Development. HIJoinCAT contains the manually observational CME events simultaneously detected by two HI cameras since 2007, and HIGeoCAT provides their kinematic properties, including the propagation directions and speeds, estimated by using the Fixed-Phi, Harmonic Mean and Self-Similar Expansion techniques. During the period of interest, HIJoinCAT lists 198 CME events. But we find that images of 3 events are missing or damaged, and therefore exclude them. Then the other 195 events are classified into three levels based on the performance of the CORAR reconstruction (see Figure 1), by visual inspection of the completeness and distortion. For the events at Level 1, the high-cc regions are disorganized and appear as scattered dots or small parts. They do not reflect the characteristics of a CME at all, but just show many pieces. At Level 2, high-cc regions are not fragmented as Level 1 events and propagate outward with time. They can be recognized as parts of a CME. In most cases, these parts belong to fronts or leading edges of CMEs. For these events, we think that the CMEs are recognized while the quality is not good enough. The well reconstructed events are classified as Level 3. In this case, the propagating high-cc regions almost cover the whole CME.

Besides, the "collinear effect", which can cause faked high-cc regions near the connecting line of two spacecraft, should be noted (Li, et al. 2018; Lyu, et al. 2020). Here, the collinear effect in the CME reconstruction is also assessed into three levels by visual inspection (see Figure 2): the events as Level 1 are severely affected by the collinear effect, resulting in the artificial high-cc region extended along the connecting line; for the events at Level 2, the collinear effect results in the presence of unreal structure, but real CME patterns can still be identified; for the events at Level 3, the reconstruction is not influenced by the effect. The numbers of the CME events in these levels are summarized in Table 1. The CME events at Level 2 or 3 of both performance and collinear effect are included in the following analysis of the optimal separation angle of the STEREO spacecraft, resulting in a total of 165 events. 156 among them have kinematic properties in HIGeoCAT fitted by STEREO-A or -B data, while 14 have results only from one spacecraft. The classification of collinear effect is used to find out CME events unsuitable for CORAR reconstruction, and the reconstruction performance is classified to determine the optimal angle.

Table 1 The numbers and proportions of events at each level

|  | Level 1 | Level 2 | Level 3 |
|---|---|---|---|
| Reconstruction performance | 10 (5.1%) | 71 (36.4%) | 114 (58.5%) |
| Collinear effect | 20 (10.2%) | 44 (22.6%) | 131 (67.2%) |

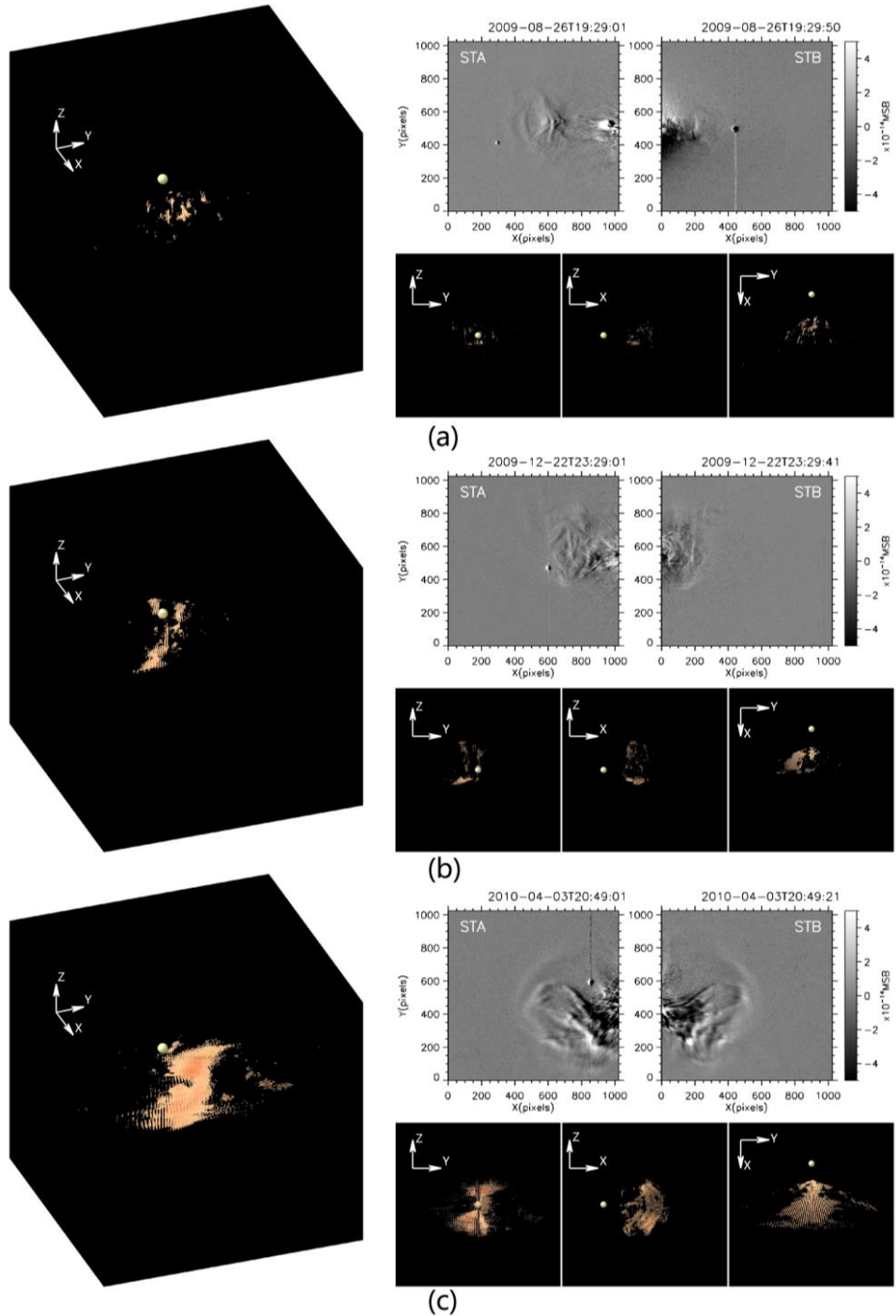

(a)

(b)

(c)

**Figure 1** Panel (a): the HI-1 images and reconstruction presentation in HEE Coordinate of 2009-08-26 CME event at Level 1 of reconstruction quality. Panel (b): 2009-12-22 CME event at Level 2 of reconstruction quality. Panel (c): 2010-04-03 CME event at Level 3 of reconstruction quality.

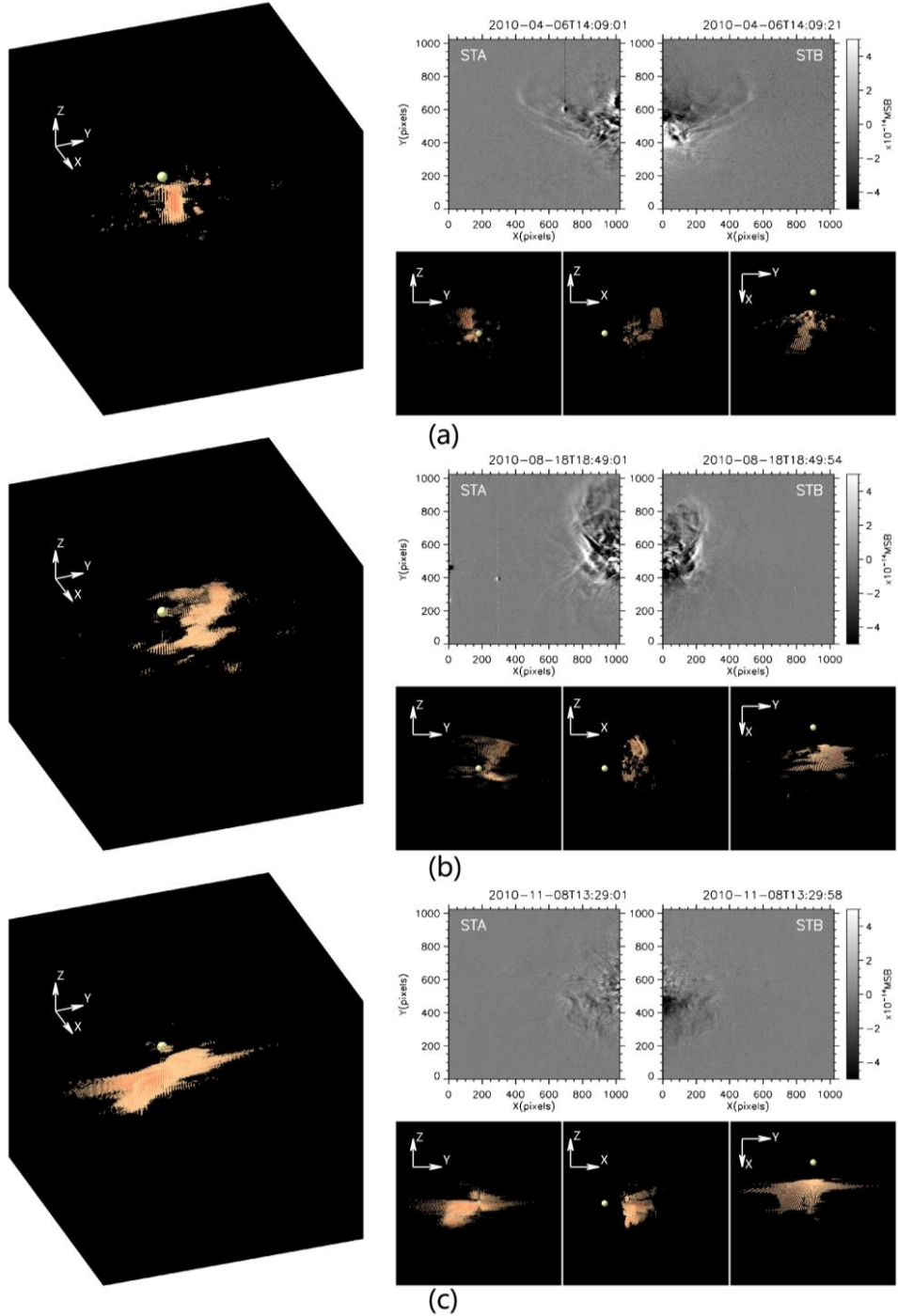

**Figure 2** Panel (a): the HI-1 images and reconstruction presentation in HEE Coordinate of 2010-04-06 CME event at Level 3 of collinear effect. Panel (b): 2010-08-18 CME event at Level 2 of collinear effect. Panel (c): 2010-11-08 CME event at Level 1 of collinear effect.

## 3. Results

We obtain the optimal stereoscopy for CORAR according to the three-dimensional reconstruction performance of CMEs in the heliosphere. Figure 3 shows event proportions at three levels of the collinear effect in Panel (a) and of the reconstruction quality in Panel (b) as a function of the separation angle, as well as the histogram of the total number of dual-view CME events for study. The errors are estimated by $\sqrt{p(1-p)/n}$, where $p$ is the proportion and $n$ is the number of events in each interval. Since the events in 2008-2009 are scarcer than those after the time period, the first two angle intervals are set to 85°- 115° and 115°-135° to increase the event number for statistical analysis, and the subsequent intervals are 10°. Note that most CME events were observed when the separation angle is near 180°. It possibly results from the increasing common space near the Thomson spheres (DeForest et al. 2013; Howard & DeForest 2012; Howard et al. 2013) of two cameras, which makes it easier to recognize the same CME from two perspectives. When the angle is larger or smaller than 180°, the lines of sight (LOS) from two observers tend to be perpendicular, observing different two-dimensional features of the same optical-thin transient. Meanwhile, the time period selected for our study belongs to the ascending phase of Solar Cycle 24, so CME frequency increases as the separation angle increases over time. The data plotted in Figure 3b excludes the CME events at Level 1 of collinear effect.

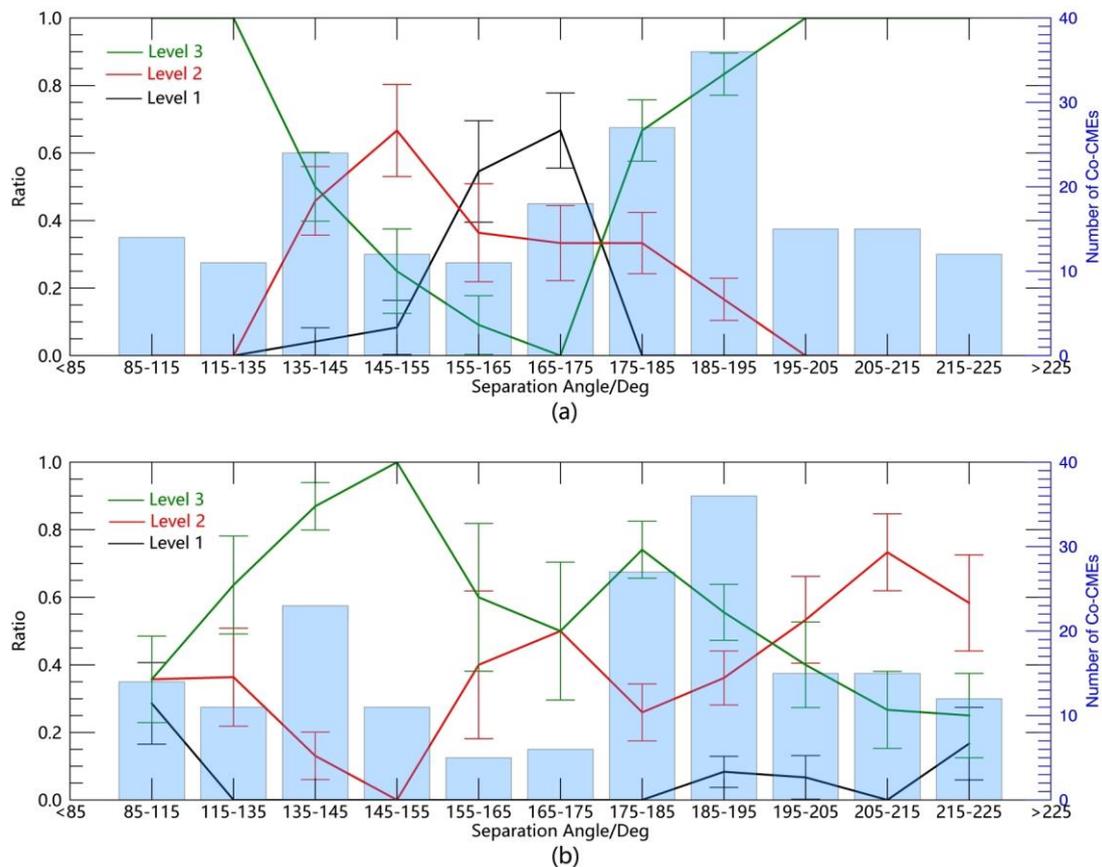

**Figure 3** Panel (a): the proportions of CME events with error bars at different levels of the collinear effect as a function of the separation angle. Events with angle of 85°-115° and 205°-225° are at Level 3. Panel (b): the proportions of CME events at different levels of reconstruction performance, without events at Level 1 of collinear effect. The number of CME events in each angle interval is plotted in light blue bars.

From Figure 3a, the proportions of CME events without collinear effect are less than 30% when the separation angle is between 145°-175°, especially more than 50% of events suffer from severe collinear effect in the 155-175° interval. Therefore, most events during this period are considered unsuitable for the CORAR reconstruction and excluded from the following analysis. There are few CME events at Level 1 of reconstruction quality in Figure 3b, especially no events between 115°-185°. It indicates that the dual-view CMEs can be recognized easily according to the CORAR reconstruction over the selected angular range. With a larger or smaller separation angle, it is less likely to leave common features of a CME on the dual images, leading to an increase of Level 1 events. For events at Level 2 and Level 3 of reconstruction performance, the proportion curves have different tendencies: the Level 3 profile reaches the maximum in 145°-155° and has the local minimum in 155°-165°, while the case for Level 2 is the opposite. More than 85% of events between 135°-155° are reconstructed well, implying the similar features of a CME left completely in the dual-perspective images. In the range of 155°-175°, the nearly parallel lines of sight from two observers bring about severe collinear effect. Although we exclude bad events, almost half of the effective CME events are at Level 2 of reconstruction quality. Around 180°, the proportion of Level 3 events rises to 0.74, and decrease monotonously as the separation angle gets larger, with more CMEs partially reconstructed. In summary, when the stereoscopic angle is $150 \pm 5°$, most CME events are well reconstructed, even if influenced to some extent by the collinear effect. On the other hand, when the stereoscopic angle is around 120°, the collinear effect is minimized with all the reconstruction quality of level 2 or 3. The compromise between the two factors suggests that the optimal separation angle of the two spacecraft should locate between 120° and 150°. This result is similar to that of our previous study (Lyu, et al. 2020), in which the separation angle of 120° is concluded as the best one.

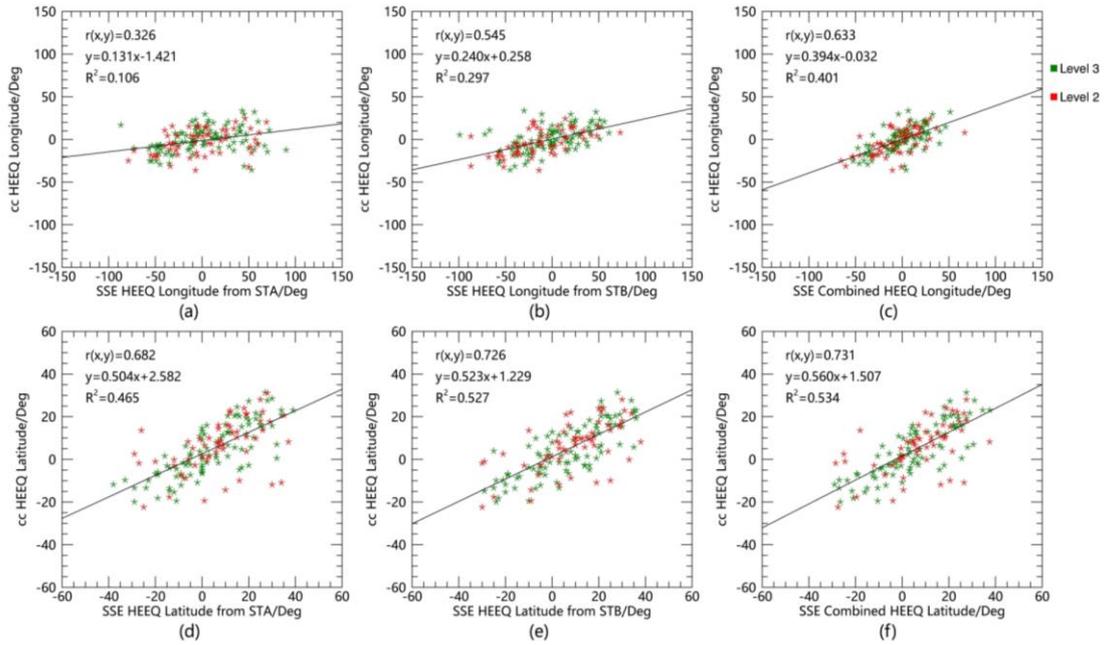

**Figure 4** The comparison of fitting longitude (Panel a-c) and latitude (Panel d-f) in the Heliocentric Earth Equatorial (HEEQ) Coordinate from CORAR method and from Self Similar Expansion (SSE) model, with fitting profiles of all CORAR results as a function of SSE results (black lines). The correlation coefficient $r$, the fitting functions as well as the coefficients of determination $R^2$ are listed. The horizontal axes represent the SSE longitude and latitude, and the vertical axes represent the direction parameters obtained by the CORAR method. SSE results are fitted from STEREO-A images in Panel (a) and (d) and from STEREO-B images in Panel (b) and (e). The SSE results in Panel (c) and (f) are mean values from STEREO A and B. The CME events at Level 3 of reconstruction quality are marked in green, and events at Level 2 in red.

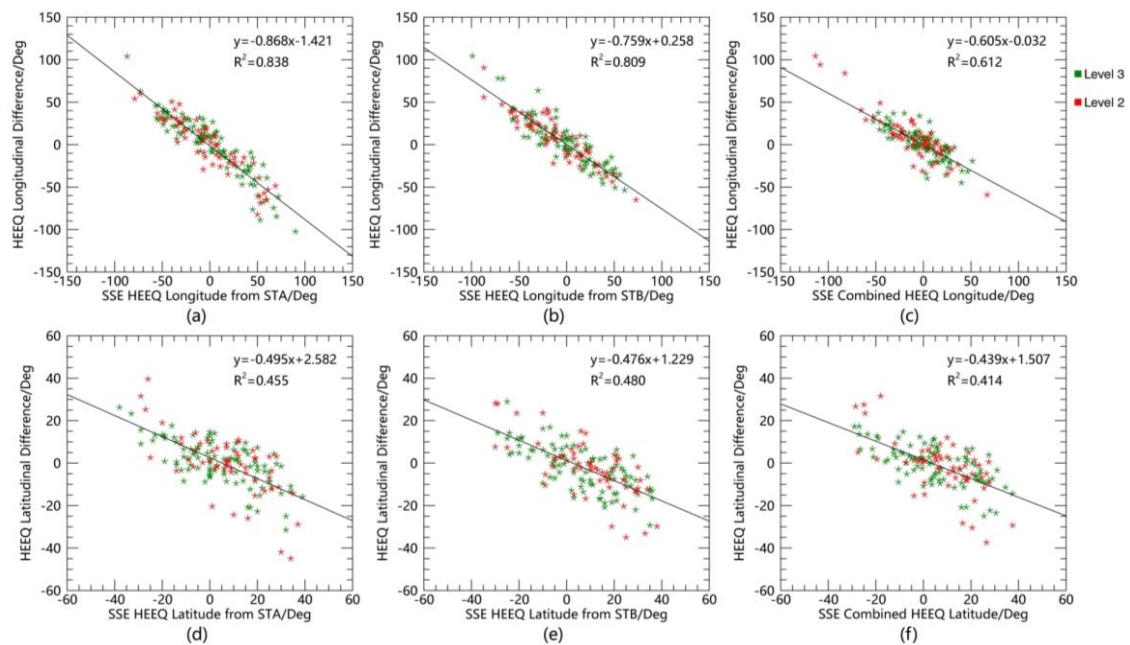

**Figure 5** The errors between CORAR and SSE longitude $\varphi_{CORAR} - \varphi_{SSE}$ (Panel a-c) and latitude $\theta_{CORAR} - \theta_{SSE}$ (Panel d-f) in HEEQ coordinate, and their fitting profiles (black lines). The fitting functions and the coefficients of determination $R^2$ are listed. The horizontal axes represent the SSE longitude and latitude, and the vertical axes represent the CORAR results minus SSE results. SSE results are fitted from STEREO-A images in Panel (a) and (d) and from STEREO-B images in Panel (b) and (e). The SSE results in Panel (c) and (f) are mean values from STEREO A and B. The CME events at Level 3 of reconstruction quality are marked in green, and events at Level 2 in red.

To discuss about the accuracy of CME location, we select effective CME events at Level 2 or Level 3 of reconstruction quality and collinear effect to track the three-dimensional trajectories of reconstructed CMEs. Assuming that CMEs move in the radial direction, we calculate the cc-weighted center of a CME, and take the average position in longitude and latitude during its propagation as the CME direction. The cc-weighted longitude $\varphi$ and latitude $\theta$ are calculated as follows

$$\varphi = \frac{\sum_i cc_i \varphi_i}{\sum_i cc_i} \qquad (1)$$

$$\theta = \frac{\sum_i cc_i \theta_i}{\sum_i cc_i} \qquad (2)$$

where $\varphi_i$ and $\theta_i$ are the latitude and longitude of any point in the high-cc region. Figure 4 compares the CME propagation directions in the Heliocentric Earth Equatorial (HEEQ) Coordinate system obtained by the CORAR method and by the self-similar expansion (SSE) fitting technique (Davies, et al. 2012) with a fixed 30° half width from HIGeoCAT. The calculated correlation coefficient $r$ larger than 0 indicates a positive correlation between CORAR and SSE parameters. Meanwhile, taking the SSE parameters as the independent variable, we can obtain linear fits with positive slopes (black lines in Figure 4), implying the prediction accuracy of the CME direction obtained by the CORAR method. If there was a perfect agreement, the slope should be unity. Compared with the single-view SSE parameters, the averages from two spacecraft may be more precise. This can partially explain the larger correlation coefficients and the slopes of fitting lines in Figure 4c and 4f than those in other panels. However, as the SSE longitude of CME events increases, the increasing magnitude of CORAR longitude is significantly smaller, revealed by the fitting slopes generally below 0.5. One possible explanation is that the SSE longitude recorded in HIGeoCAT is fitted for the CME apex instead of the whole CME, and fitting errors may come from the single-view fitting methods and the reconstruction CORAR method. Nevertheless, this phenomenon possibly reflects a clear deviation toward the meridian plane in longitude for the reconstructed CMEs by CORAR. It is more intuitive in Figure 5, which shows the difference between CORAR and SSE values. It is apparent that the deviation in longitude increase with the absolute longitude.

In the direction of latitude, the correlations between CORAR and SSE values are relatively stronger, and the fitting slopes in Figure 4d-4f are closer to 1 than those in longitude, but there are deviations toward the equatorial plane, similar to the longitudinal deviations (see Figure 5d-5f). The deviations may be related to the unphysical structures caused by the intrinsic defects of triangulation method. The specific analysis will be discussed in the subsequent section. A large uncertainty may exist when using CORAR methods to locate CMEs with different scales and morphologies in the same propagating direction, because our CORAR method does not incorporate any morphological assumptions and takes the whole irregular structures of reconstructed CMEs for tracking trajectories. The HIGeoCAT catalogue also provides the CME kinetic parameters obtained from other two fitting models: the Fixed-Phi (FP) model and Harmonic-Mean (HM) model, which are considered as the extreme cases of the SSE with a half width of 0° and 90°, respectively (Davies, et al. 2012; Moestl & Davies 2013). Since the comparison between values from the two fitting models and from our CORAR method is similar to the analysis above, the details are not described here.

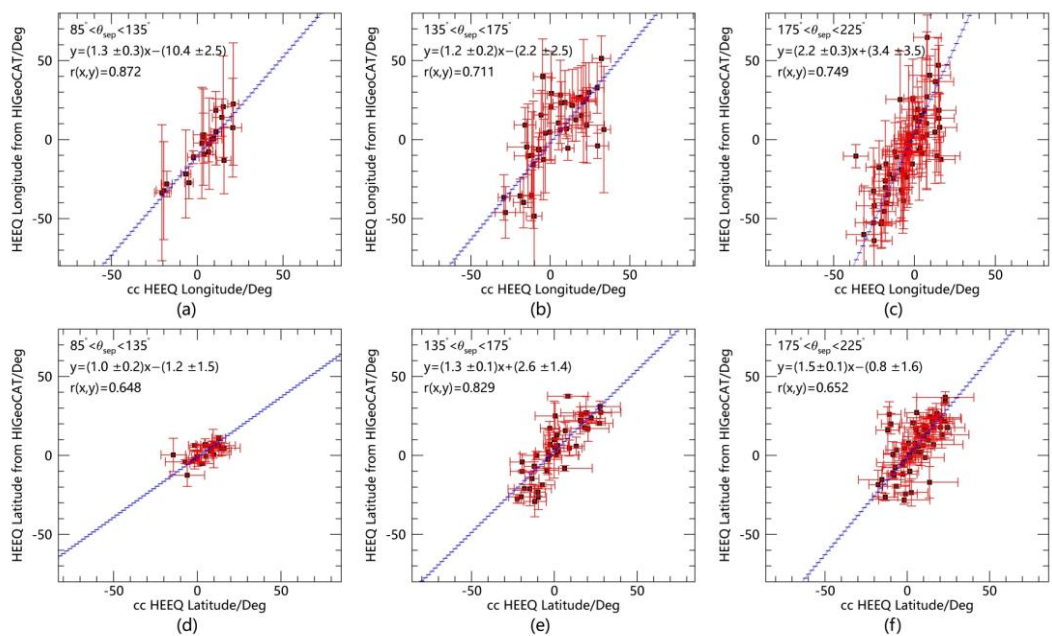

**Figure 6** The fitting lines for correction of CME directions from CORAR in HEEQ longitude (Panel a-c) and latitude (Panel d-f) with the separation angle of 85°-135° (Panel a and d), 135°-175° (Panel b and e), and 175°-225° (Panel c and f).

Based on above comparative analysis, we try to correct the CME propagation directions obtained by CORAR method. Since the deviation may depend on the separation angle of the spacecraft, we use piecewise functions to make the correction as shown in Figure 6. Note that when the separation angle ranges from 135° to 175° approximately, the connecting line of two spacecraft basically exists in the common HI-1 field of view, so that the collinear effect is serious (Figure 3a). Meanwhile, this angular interval also contains the most suitable separation angle for good reconstruction. Therefore, we divide the function interval into three: 85°-135°, 135°-175°, and 175°-225°. We take the CORAR parameters as the independent variable to linearly fit the average values from the single-view fitting models in HIGeoCAT, and get the empirical formulas for the correction of CORAR values as follows

$$\varphi = \begin{cases} (1.3 \pm 0.3)\varphi_{CORAR} + (-10.4 \pm 2.5), & 85° < \theta_{sep} < 135° \\ (1.2 \pm 0.2)\varphi_{CORAR} + (-2.2 \pm 2.5), & 135° < \theta_{sep} < 175° \\ (2.2 \pm 0.3)\varphi_{CORAR} + (3.4 \pm 3.5), & 175° < \theta_{sep} < 225° \end{cases} \quad (3)$$

$$\theta = \begin{cases} (1.0 \pm 0.2)\theta_{CORAR} + (-1.2 \pm 1.5), & 85° < \theta_{sep} < 135° \\ (1.3 \pm 0.1)\theta_{CORAR} + (2.6 \pm 1.4), & 135° < \theta_{sep} < 175° \\ (1.5 \pm 0.1)\theta_{CORAR} + (-0.8 \pm 1.6), & 175° < \theta_{sep} < 225° \end{cases} \quad (4)$$

$\theta_{sep}$ is the separation angle, and $\varphi_{CORAR}$ ($\theta_{CORAR}$), $\varphi$ ($\theta$) are initial and corrected HEEQ longitude (latitude) obtained by CORAR method, respectively. The fitting method is different from those in Figure 4 and 5, considering both errors from CORAR and modelling results. The CORAR errors measure the possible deviations of reconstructed CMEs from the average propagation directions, and the errors of modelling results are standard deviations of six values from three single-view techniques by fitting STEREO-A and –B images. The slopes of correction lines for HEEQ longitude and latitude are slightly larger than 1 when $\theta_{sep} < 175°$. Above this range, the slopes of the correction formulas are apparently larger than 1, especially in longitude. It may result from the increase of events at Level 2 of reconstruction quality, with their partial structures in smaller absolute longitude and latitude reconstructed better. It indicates a separation angle larger than 180° may not be suitable for CME reconstruction by CORAR method.

## 4. Discussion

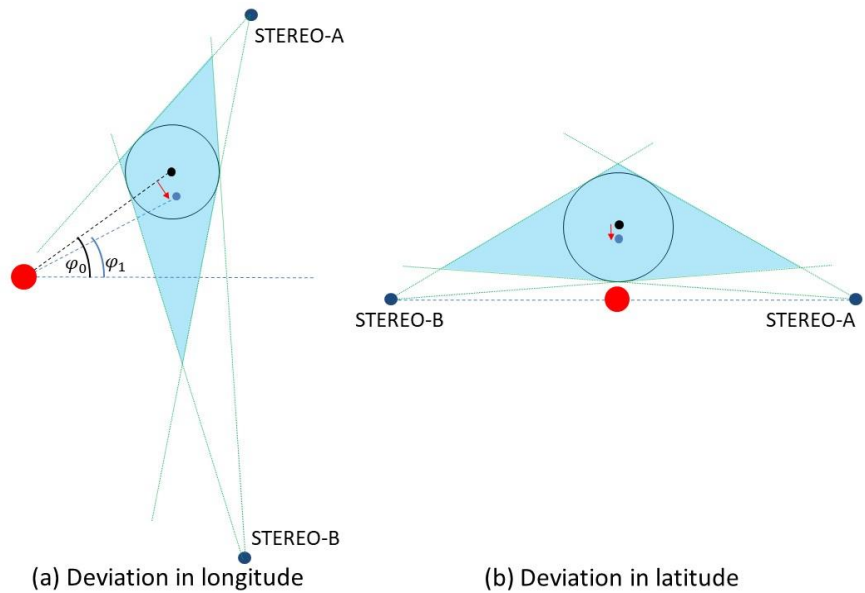

(a) Deviation in longitude  (b) Deviation in latitude

**Figure 7** The principle of systematic errors modelled by spherical shell in longitude

and in latitude. Panel (a): the model observed along the Z axis in HEE coordinate. Panel (b): the model observed along the Sun-Earth line. The red ball represents the Sun and two blue balls mark the two STEREO spacecraft. The black and blue point represent the center of the CME model and the blue area respectively.

Mierla, et al. (2010) demonstrates that there are many factors affecting the reconstruction quality and location accuracy for three-dimensional reconstruction methods, which can be roughly divided into observational errors and methodical errors. For the CORAR method, the possible errors may come from three aspects. First, the similar features of the same 3D structure cannot be observed from dual perspectives, so the reconstruction cannot be performed. Second, the similar patterns that do not belong to the same transients are mistaken for real structures, thus the reconstruction contains unreal transients. Third, similar features of the same structure can be observed in dual-view images, but fails to be reconstructed completely due to factors like too large sizes of the sampling box or grid cells. The first type is common for reconstruction methods based on multi-view observations, especially when the separation angle of spacecraft is close to 90°. From different LOSs for Thomson scattering integral, the same transient may have completely different patterns, so that the errors are difficult to deal with. Considering that the characteristics of large-scale transients can be sufficiently recognized in details based on the high resolution of HI-1 images and the control parameters can be adjusted, the third type of errors is easy to control. As for the second type of errors, it may be theoretically estimated and further corrected.

To analyze the second errors, we apply a simple spherical model to simulate the 195 CME events, as shown in Figure 7. The SSE fitting parameters of the corresponding CME event are assumed as the actual propagating direction of the CME for the spherical model. The radial distance from the Sun is determined by the instantaneous position of the reconstructed CME from CORAR method, and the spherical radius is determined by the half-width of the reconstructed CME in latitude. The blue area is theoretically the space containing all possibly reconstructed structures in the common field of view, assuming a homogenetic distribution. To estimate the influence of these false structures, we calculate the cc-weighted center of this area so that its deviation from the center of the initial model can be obtained, as shown in Figure 8. The existence of these false structures causes the calculated propagation directions of CMEs to shift toward the meridian plane in longitude and the equatorial plane in latitude, which is more serious for CME events with larger characteristic scales. Especially, CME events with a half width larger than 30° may produce an angular deviation of more than 10° in longitude and latitude. Furthermore, we subtract the aforementioned errors from the CORAR parameters for each CME event in Figure 4, which are shown in Figure 9. After subtracting the calculated systematic errors, the slopes of fitting lines for CORAR latitude and longitude approach to unity, and the correlation coefficient $r$ and the statistical measure $R^2$ generally increase. In particular, Figure 9c and 9f show fitting lines with slopes near 1 and a positive correlation larger than 0.8 for average SSE parameters. However, the corrected fitting straight lines are still less than 1. Although the single-view fitting parameters cannot be regarded as the true values of CME propagating directions, the theoretical deviation cannot fully explain the deviation from the CORAR method in Figure 4 and 5. Remind that the blue space in Figure 5 is an extremely ideal case. In reality, the reconstructed transients may only occupy small parts of the blue space, or even scattered points for CME events at Level 1 of reconstruction performance. The spherical model is too simple to fit the complex solar wind structures. In addition, other solar wind transients existing in the observation images, such as blobs or successive CMEs, may generate structures outside the blue space. Finally, this error analysis requires that the cc distribution, which can measure the authenticity of the reconstructed inhomogeneous transient, is approximately uniform, but it is not true for most CME events. More work is needed to analyze the source of the deviation in further detail for complete error correction.

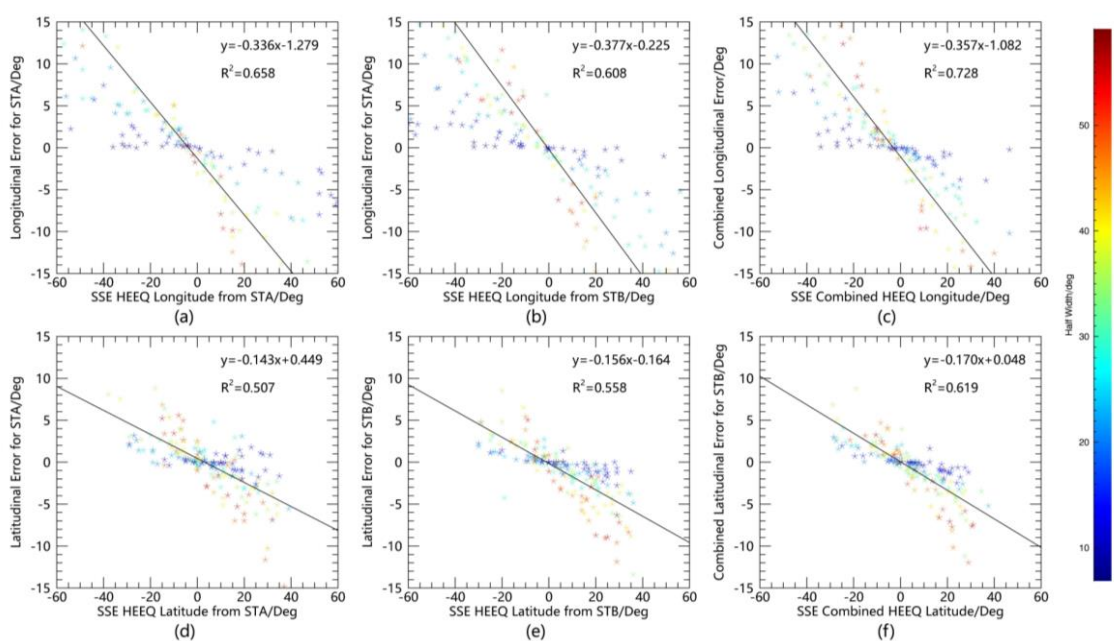

**Figure 8** The systematic errors in HEEQ longitude (Panel a-c) and latitude (Panel d-f) calculated by simple spherical model for the CME events, with the fitting lines in black. The horizontal axes represent SSE-fitting longitude and latitude, and the vertical axes represent the theoretical errors. SSE results are fitted from STEREO-A data in Panel (a) and (d) and from STEREO-B data in Panel (b) and (e). The SSE results in Panel (c) and (f) are mean values of that from STEREO A and B. The color bar represents the latitudinal half width of CME events.

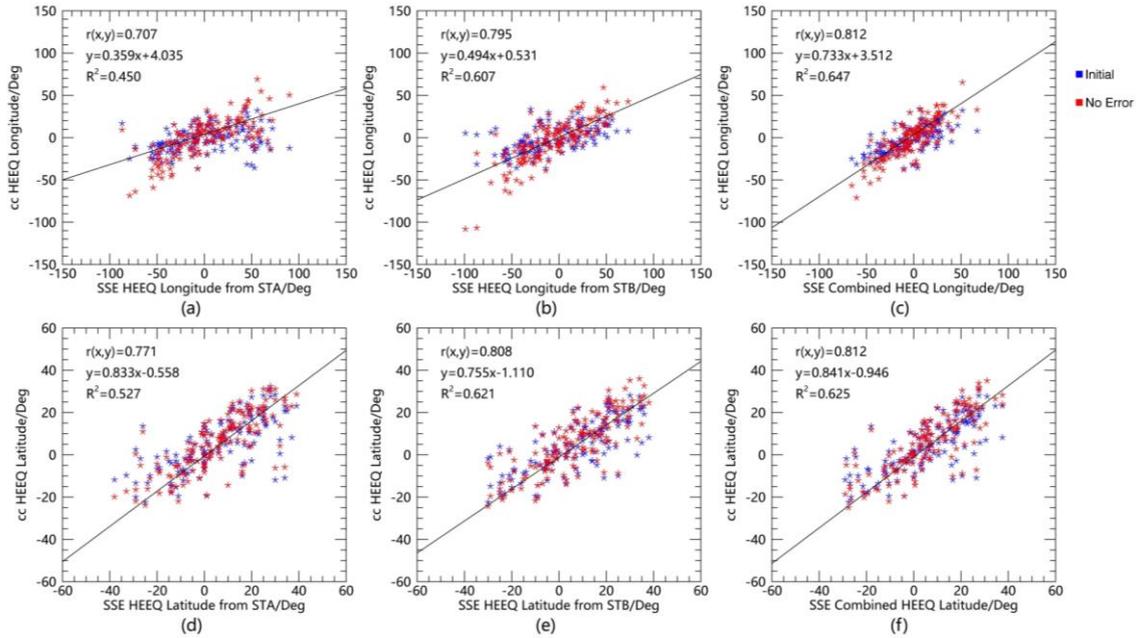

**Figure 9** The CORAR-calculated longitude (Panel a-c) and latitude (Panel d-f) corrected by systematic error in Figure 8, as well as their fitting lines (black lines). The correlation coefficient $r$, the fitting functions as well as the coefficients of determination $R^2$ are listed. The horizontal axes represent SSE-fitting results, and the vertical axes represent the CORAR results minus errors (blue) and those without error correction (red). SSE results are fitted from STEREO-A data in Panel (a) and (d) and from STEREO-B data in Panel (b) and (e). The SSE results in Panel (c) and (f) are mean values from STEREO A and B.

## 5. Conclusion

In our previous work (Lyu, et al. 2020), we studied the 3D reconstruction of small-scale blobs by the CORAR method, and discussed the optimal stereoscopic angle for the Solar Ring mission (WANG et al. 2020a; Wang, et al. 2020b). In this study, we use the CORAR method to reconstruct the HI-1 dual-perspective images from December 2008 to February 2012, obtain the three-dimensional cc distribution to study the reconstruction performance for CME events in the heliosphere, and calculate the propagation direction of transients. The CME events for study come from the HIJoinCAT catalogue in the HICATS project, which consists of dual-view CME events, so that the theoretical three-dimensional reconstruction is ensured.

We find that the influence of the collinear effect clashes with the trend for the variation of the reconstruction quality as the separation angle increases: when the stereoscopic angle rises from 90° to 180°, the CME patterns in the dual-perspective images may be more similar, suitable for CORAR reconstruction; at the same time, the spacecraft appear in the FOV of each other, bringing difficulty in locating the accurate position of CMEs in the direction of the connecting line and resulting in large-scale fake structures. Among the angular intervals, 120°-150° is considered as suitable for CME reconstruction. Within 145°-155°, some CME events are influenced by the collinear effect, but most cases are well reconstructed. When the separation angle is smaller, the proportion of CME events with good reconstruction quality decreases, while the collinear effect weakens until 115°. Below this range, CMEs may fail to be reconstructed. As the separation angle gets larger than 155°, the influence of the collinear effect becomes more serious so that a considerable amount of CME events are not reconstructed well.

We also calculate the propagating directions of CMEs in HEEQ coordinate and compare them with single-view fitting results from the HIGeoCAT catalogue, so as to correct the CORAR results by fitted empirical formulas in different angle intervals. We prove that the CORAR results are positively correlated with the SSE-fitting results, while apparent deviations toward the meridian plane in longitude and smaller shifts toward the equatorial plane in latitude exist. We speculate that the reason for the deviations is mainly due to the unreal structure produced by the similar characteristics of different CME patterns. For further discussion, we use a simple spherical model to calculate the theoretical error for each CME event. It can partially explain the sources of the deviation and improve the CORAR prediction of CME directions. To completely analyze and correct this deviation, distinguish the difference between the real and unreal structures, and realize absolute reconstruction for various solar wind inhomogeneous transients, further research is needed in the future.

Our work supports the spacecraft scheme of the Solar Ring mission (Wang, et al. 2020b). In this plan, the separation angles among the six spacecraft orbiting the Sun have the values of 120° and 150°. According to our studies, the 120° scheme is suitable for small-scale transients, and the 150° scheme can work for large-scale CMEs. We hope our work can be helpful for the design of this mission as well as other mission concepts for solar wind observations in the future.

**Acknowledgements**    We acknowledge the use of the data from STEREO/SECCHI, which are produced by a consortium of RAL (UK), NRL (USA), LMSAL (USA), GSFC (USA), MPS (Germany), CSL (Belgium), IOTA (France), and IAS (France). The SECCHI data can be found in the STEREO Science Center (https://stereo-ssc.nascom.nasa.gov/data/ins_data/secchi_hi/L2/). We also acknowledge the service provided by the data repository at "National Space Science Data Center, National Science & Technology Infrastructure of China. (http://www.nssdc.ac.cn/eng/)", where our data products are stored. This work is supported by the Strategic Priority Programs of the Chinese Academy of Sciences (XDA15017300 and XDB41000000), the National Natural Science Foundation of China (41774178, 41761134088, 41804161 and 42074222) and the fundamental research funds for the central universities (WK2080000077). Y.W. is particularly grateful to the support of the Tencent Foundation.